\documentclass[12pt]{article}
\usepackage{epsf}

\newcommand{\bee}{\begin{equation}}
\newcommand{\ee}{\end{equation}}
\newcommand{\bea}{\begin{eqnarray}}
\newcommand{\eea}{\end{eqnarray}}
\newcommand{\R}{\rm I\kern-.2emR}
\newcommand{\C}{\rm \kern.25em\vrule height1.4ex
depth-.12ex width.06em\kern-.31em C}
\newcommand{\N}{{\rm I\kern-.16em N}}
\newcommand{\Z}{{\rm Z\kern-.35em Z}}
                                                      
\begin{document}                                                                
\begin{flushright}
MPI-PhT/2001-46\\
\end{flushright}
\bigskip\bigskip
\begin{center}
\Huge{Some more remarks on the Witten-Veneziano formula for the $\eta'$
mass}
\end{center}
\centerline{\it Erhard Seiler}
\centerline{\it Max-Planck-Institut f\"ur Physik}
\centerline{\it (Werner-Heisenberg-Institut)}
\centerline{\it F\"ohringer Ring 6, 80805 Munich, Germany}
\centerline{\it e-mail: ehs@mppmu.mpg.de}
\bigskip \nopagebreak
\begin{abstract}  \noindent
We discuss some subtleties in connection with the new attempts to
provide a firm basis for ths Witten-Veneziano formula.
\end{abstract}
\vskip2mm
\section{Introduction}

More than twenty years ago Witten \cite {witten} and Veneziano
\cite{veneziano} proposed a formula connecting the mass of the
$\eta'$ meson to the quenched topological susceptibility 
$\chi_t^{\rm qu}$ of QCD. This formula takes its simplest form in
the chiral limit:
\bee
m^2_{\eta'}={2N_f\over F_\pi^2} \chi_t^{\rm qu}
\label{WV}
\ee
Recently Giusti, Rossi, Testa and Veneziano \cite{grtv} tried to put 
the old arguments on a firmer basis by starting with a well-defined
lattice version, taking advantage of the recent progress in the
understanding of chiral lattice fermions (see for instance the review
\cite{ferenc} and references therein).

The crucial input for all derivations is the anomalous Ward identity for
the $U(1)$ axial current, which leads to the vanishing of the topological
susceptibility in the full theory with dynamical fermions in the chiral 
limit:
\bee
\chi_t^{\rm full}=0
\label{sumrule}
\ee

But the interpretation of both equations is tricky, as was pointed out 
long ago in \cite{ss}; there it was also observed that Witten's
original arguments would require cancellation of terms of equal signs 
against each other. (The 1987 paper remained unpublished, but a scanned 
version is available from KEK via Spires). Here I will try to explain 
more clearly the main point of that old paper, correct some imprecisions 
and discuss its implications for the recent attempts.

The conclusion will be that the WV formula can be given an interpretation
that makes it true, but that it is ambiguous as it stands. The recent
lattice approaches offer the prospect of eliminating this ambiguity
and providing a reliable foundation for the formula. It will also become
clear that short-distance fluctuations of the topological density
do play a crucial role.

\section{`Axiomatic' considerations}

There is a rather close analogy between $4D$ QCD and $2D$ (multiflavor) 
QED; the latter model, being exactly soluble, provides therefore a good
testing ground for the considerations connecting mass generation in the
flavor-neutral pseudoscalar channel to the topological susceptibility. 
We will at first formulate `axiomatically' the conditions which the 
topological charge density operator should fulfil in a continuum quantum
field theory; in this general discussion the $4D$ and $2D$ models can be
handled together. In the case of ${\rm QED_2}$ the `axioms' are true statements 
that hold in the explicit constructions. The situation for ${\rm QCD_4}$ 
is different, however, since the Millenium Prize problem \cite{clay} of 
constructing this theory with or without fermions has not yet been 
solved. Here we assume here that such a theory exists and that its 
gauge invariant fields satisfy the Wightman axioms or, after 
continuation to the euclidean world, the Osterwalder-Schrader axioms 
(see for instance \cite {gj}).

Initially we do not have to distinguish between the quenched (no dynamical
fermions) and the full models. The field of interest is the topological
density, given in ${\rm QED_2}$ by
\bee
q(x)={e\over 2\pi} F_{01}(x)
\ee
and in ${\rm QCD_4}$ (formally)  by
\bee
q(x)={g^2\over 32 \pi^2} \rm{tr}\  F_{\mu\nu}(x) \tilde F_{\mu\nu}(x)
\ee
where $F$ is the Yang-Mills field strength tensor and $\tilde F$ its
dual. The crucial point is to notice that $q$ is odd under time
reflections and therefore satisfies an unusual form of reflection 
positivity (RP), which for its 2-point function reads
\bee
G(x)=\langle q(x)q(0)\rangle\leq 0 \ \ {\rm for}\ \ x \neq 0.
\label{RP}
\ee
Another way of saying this is that the Euclidean field $q(x)$ 
corresponds to an antihermitian field operator because it contains
one time derivative. This form of RP was stressed in \cite{ss}
and later in \cite{vicari}.

The topological susceptibility $\chi_t$ is supposed to be the integral 
$\int G(x){\rm d}x$. Two questions arise immediately: \\
\noindent
(1) {\it Is $G(x)$ integrable over the whole space?}\\
\noindent
(2) {\it How can $\chi_t$ be positive, as required by eq. (\ref{WV}), if
the integrand is negative?}

The answers to these questions are closely related. For (1) one would
expect a negative answer: in ${\rm QED_2}$ $q(x)$ seems to be a dimension
2 field, whereas in ${\rm QCD_4}$  one has to expect that it has
dimension 4; in both cases $G(x)$ should not be expected to be
integrable at short distances.

On closer inspection, it actually turns out that in ${\rm QED_2}$
there is a lucky coincidence: the coupling constant $e$ has the dimension
of a mass and the topological density turns out to be proportional
to $e$ times a dimension 0 field plus some white noise producing
a contact term (see \cite{s}).

This kind of accident cannot be expected in ${\rm QCD_4}$. To give
the space-time integral of $G(x)$ meaning, counterterms
concentrated at $x=0$, i.e. divergent contact terms are needed (see 
for instance \cite{gelfand}).  The answer to question (2) is then that
with a suitable choice of those contact terms one can indeed make 
$\chi_t$ nonnegative. The validity of formulae like eq. (\ref{WV}) and
(\ref{sumrule}) thus depends crucially on the right choice of contact
terms.

We will now discuss the two cases ${\rm QED_2}$ and ${\rm QCD_4}$ 
separately in a little more detail. 

\section {${\rm \bf QED_2}$}

This case has been discussed for one flavor in \cite{ss} and for $N_f$
flavors in \cite{gs}. The construction employed there fixes the
possible contact terms (which are finite in this case). We cite from the
latter reference the result
\bee
\hat G(p)={e^2\over 4\pi^2} \left(1-{e^2(N_f/\pi)\over p^2+e^2(N_f/\pi)}
\right)
\label{QED2}
\ee
for full ${\rm QED_2}$ with $N_f$ dynamical fermions; the quenched  
correlation is obtained by setting $N_f=0$ and is just the pure contact
term contained in eq.(\ref{QED2}).

As was discussed in \cite{ss,gs}, this construction and the choice of
contact terms inherent in it make the formula eq. (\ref{WV}) true, provided
$F_\pi$ is interpreted appropriately. As remarked before, it is a 
special feature of this two-dimensional model (related to the fact that 
the charge has the dimension of a mass) that $G(x)$ is, aside from a 
$\delta$-function, an integrable function. Correspondingly its Fourier 
transform satisfies a dispersion relation (= K\"allen-Lehmann 
representation) of the form
\bee
\hat G(p)=c-\int_0^\infty {\rm d}t {\rho(t)\over t+p^2}
\ee
where the constant $c$ is, up to some trivial numerical factor, equal 
to the quenched topological susceptibility $\chi_t^{\rm qu}$ and the
spectral density $\rho$ is a $\delta$-function.

\section {${\rm \bf QCD_4}$}

By dimensional analysis and tree level perturbation theory $q(x)$ is 
expected to be a dimension 4 field and hence, up to possible logarithms
\bee
G(x)=O({1\over |x|^8}) \ \ {\rm for}\ \  x\to 0
\ee
The Wightman axioms, which are assumed to hold for $q(x)$, guarantee that
in the Euclidean world $G(x)$ is an analytic function for $x\neq 0$.
Before we can talk about the Fourier transform $\hat G(p)$ of $G(x)$ or
the topological susceptibility, we have to promote $G(x)$ to a 
distribution, and this means prescribing certain formally divergent 
contact terms. Mathematically the procedure goes as follows: $G(x)$
can already be smeared with test functions that vanish to sufficiently
high order at the origin, i.e. this smearing defines a linear functional
on a certain subspace of the test function space. To extend this linear
functional to all test functions in a way that is consistent with
euclidean invariance requires the choice of 3 free parameters,
corresponding to counterterms of form
\bee
 c_1\delta(x),\ \ c_2 \Delta \delta(x), \ \ c_3 \Delta^2 \delta(x)
\ee
Once the extension has been fixed, the distribution $G$ can be
Fourier transformed according to the rules for distributions (see for
instance \cite{gelfand}).

Since neither in Yang-Mills theory nor in full QCD we expect the 
presence of a massless pseudscalar particle with the quantum numbers
of $q(x)$, we will make the further assumption that $G(x)$ decays
exponentially at large $|x|$ 
\bee
G(x)=O\left(\exp\left(-m|x|\right)\right)\ \ {\rm for}\  x\to\infty
\ee
This has the consequence that the Fourier transform $\hat G(p)$ is 
analytic in a neighborhood of real momenta; reinterpreted as a function 
of $p^2$ it is analytic near the real axis except for a cut from $-\infty$
to $-m^2$.  At large momenta $p$ $\hat G(p)$  grows like $O(|p|^4)$ up to
some possible logarithms. By the K\"all\'en-Lehmann representation (which
actually follows from RP and euclidean invariance) we obtain the subtracted 
dispersion relation stated in \cite{grtv}
\bee
\hat G(p)=a_1+a_2p^2+a_3(p^2)^2+(p^2)^3\int_{m^2}^\infty 
\rho(t){1\over(t+p^2)t^3} dt
\label{disp}
\ee
where the constants $a_i$ are proportional to the free parameters $c_i$
and $\rho(t)dt$ is a positive measure, growing at most like $t^2$ for
$t\to\infty$. 

It is obvious from this discussion that the `topological susceptibility',
\bee
\chi_t\equiv (2\pi)^4 \hat G(0)
\ee
does not have any unambiguous meaning, be it in full or quenched QCD. 
In full QCD in the chiral limit one postulates
\bee
\chi_t^{\rm full}=0,
\ee
based on the anomalous Ward identity and the absence of zero mass
particles. This equation can clearly be made true by simply
putting $a_1=0$. It is also clear that eq. (\ref{WV}) can likewise be 
made true by a suitable choice of the constant $a_1$ for the quenched
case, but that way the formula would of course not have any predictive
value.

The authors of \cite{grtv} propose to derive the WV formula from
eq.(\ref{disp}) by first sending the parameter $u\equiv N_f/N_c$ to zero
and then going to $p=0$. A crucial assumption is that for $u\to 0$ at
fixed $p\neq 0$ the left hand side goes to the quenched value 
$\hat G(p)^{qu}$.
The right hand side is treated by an expansion in powers of
$u/p^2$ followed by sending $p\to 0$. This is a dangerous procedure,
because truncating such an expansion at order $(u/p^2)^k$ leaves an
error term $O((u/p^2)^{k+1})$, and therefore sending $p\to 0$ termwise in
the expansion is not justifiable. This problem can, however, be
circumvented by rewriting the dispersion relation eq. (\ref{disp}) in the
form
\bee
\hat G(p)=b_1+b_2p^2+b_3(p^2)^2-{R^2\over p^2+m^2_{\eta'}}+
(p^2)^3\int_{m^2}^\infty \sigma(t){1\over(t+p^2)t^3} dt   
\label{disp2}
\ee
where we have separated the contribution of the $\eta'$ meson which is
expected to dominate the dispersive integral. Simply putting $p=0$
in this equation, one arrives at the relation
\bee
b_1={R^2\over m^2_{\eta'}}.
\label{WV'}
\ee
By standard arguments one derives from this relation a WV-like formula, 
in which, however,
the contact term $b_1$ takes the place of $\chi_t^{\rm qu}$. This was
essentially the proposal made in \cite{ss} (where, however, an
unsubtracted dispersion relation was used, which is only justified {\it after} 
approximating the spectral density $\rho$ by a $\delta$-function at the 
$\eta'$ mass; see also \cite{wipf}).

This latter identification of $b_1$ with the quenched topological 
susceptibility can be based on the following reasoning, following the  
route taken by \cite {grtv}:
We also first send the parameter $u=N_f/N_c$ to zero at fixed 
$p$ and accept the assumption of \cite {grtv} that this 
corresponds to quenching on the left hand side of eq. (\ref{disp2}); 
on the right hand side, assuming with \cite{grtv} that both $m^2_{\eta'}$
and $R^2$ are $O(u)$, after taking the second limit $p\to 0$, one 
obtains just $b_1$; thus one concludes 
\bee
\chi_t^{\rm qu}=b_1,
\ee
or, using eq. (\ref{WV'})
\bee
\chi_t^{\rm qu}={R^2\over m^2_{\eta'}},
\ee
which now leads by the usual arguments to the WV equation (\ref{WV}) in
its standard form.

But the  fact remains that without suitably fixing the contact terms,
the WV relation does not hold, and, as the discussion above shows, the
quenched topological susceptibility is in fact {\it equal} to the contact
term $b_1$, similar to the situations in ${\rm QED_2}$. To put more 
meaning into the WV relation, a self-contained lattice derivation is 
certainly desirable, and the paper \cite{grtv} takes some important steps in
that direction. We will make some comments about this in the next section.

The statement that $\chi_t$ is defined only up to a free parameter
and could have either sign, seems to clash with the `obvious' identity
\bee
\chi_t=\lim_{V\to\infty} {1\over V} \langle Q_V^2 \rangle
\ee
with $Q_V=\int_V q(x){\rm d}x$,
which seems to show that manifestly that $\chi_t\geq 0$. But this argument
is too naive. A harmless point is that a sharp volume cutoff
as in $Q_V$ is not allowed due to the singular nature of the correlators
of $q(x)$. This can easily be fixed by replacing the quantity $Q_V$ by
$Q(f_V)$ where $f_V$ is a smooth approximation the characteristic
function of the volume $V$. But one can still not conclude that
$\langle Q(f_V)^2 \rangle\geq 0$, because there is no physical principle
that restricts the free parameters $c_1,c_2, c_3$. More generally, there
is no physical principle requiring that the continuum correlation 
functions are moments of a positive measure and that the symbol 
$\langle\  .\ \rangle$
used to denote euclidean expections really means an expectation value in
the probabilistic sense.  Of course if is possible to choose
$a_1\geq 0$; the  dispersion relation then guarantees that
$\chi_t=1/(2\pi)^4\hat G(0)\geq 0$. But again everything depends
entirely on the choice of the contact term, which does not have any 
intrinsic physical meaning.

\section{Lattice versions of the WV formula}

There have been earlier attempts to derive lattice versions of WV like
formulae (\cite{smitvink}), but the important progress that has taken
place in the contruction of chiral lattice fermions (see for instance
\cite{ferenc} and references therein) suggested a new attack on the
problem using Ginsparg-Wilson (GW) fermions and this is what the
authors of \cite{grtv} proposed to do.

Everything is now well defined and in the absence of any vacuum
angle $\theta$ one really has a positive measure determining the 
euclidean expectation values (Nelson-Symanzik positivity holds). So 
in this framework it is simply a fact 
that
\bee
\chi_t^\#\geq 0
\ee
for $\#={\rm qu}$ or ${\rm full}$, and the good chiral properties of 
the GW fermions assure that the anomalous Ward identity holds and hence
\bee
\chi_t^{\rm full}=0  .
\ee
The arguments sketched for the continuum depend on dispersion relations
that are not valid in this form on the lattice. But if one assumes that 
they hold up to some corrections that disappear in the continuum
limit, one obtains a lattice derivation of the WV formula which now
has an unambiguous meaning (at least once one has settled on a definite
solution of the Ginsparg-Wilson relation) and hopefully has a finite
continuum limit on both sides. So there is a good chance that the work of
\cite{grtv} can be the starting point for a solid foundation of the WV
formula.

It would be very interesting to study the approach to the continuum
of the quenched 2- point function of the topological charge density in
this GW framework and see how the subtleties discussed above emerge.
Even though RP does not hold for GW fermions before taking the continuum 
limit, it should (hopefully) become valid in this limit. So one should 
expect that in this framework the correlator of the lattice
version of $q(x)$ is negative, except at distances of a few lattice 
spacings, and one should see the emergence of a divergent contact term.

These phenomena have been studied in some detail in two-dimensional
spin models: by Balog and Niedermaier \cite{bn} in the $2D$ $O(3)$ model
and by Vicari \cite{vicari} in the $2D$ $\C P^{N-1}$ model in the $N\to\infty$
limit. In this work it can be seen clearly that the correlator is 
negative, in accordance with reflection positivity, except at coinciding 
points, where the compensating contact term emerges. An analogous study
for the case of ${\rm QCD_4}$, especially with the definition of the
topological density suggested by \cite{grtv} might be elucidating. 

The main conclusion of this discussion is: the WV formula is ambiguous
as it stands, and its truth depends strongly on the right choice of
contact terms. If one starts from the lattice, it therefore all depends
on the right treatment of the short distance fluctuations. The GW
framework offers some hope for a self-contained lattice derivation 
and the anomalous Ward identity suggests the right choice of the 
topological density with the right short distance fluctuations.

The author is grateful to P. Weisz for discussions.

\end{document}